# Nonstandard Lorentz-Einstein transformations

Bernhard Rothenstein[1] and Stefan Popescu[2]
1) Politehnica University of Timisoara, Physics Department,
Timisoara, Romania brothenstein@gmail.com
2) Siemens AG, Erlangen, Germany stefan.popescu@siemens.com

**Abstract.** *The **standard** Lorentz transformations establish a relationship between the space-time coordinates of the same event when detected from two inertial reference frames I and I' in the standard arrangement. This event is characterized by the space-time coordinates $E(x,t_E)$ and $E'(x',t'_E)$, $t_E$ and $t'_E$ representing the readings of the standard synchronized clocks $C(x)$ and $C'(x')$ located in the two frames where the event takes place. We obtain the **nonstandard** Lorentz transformations establishing a "physically" correct relationship between the readings of the standard synchronized clocks and the readings of other clocks $(t_a, t'_a)$ of the same inertial reference frames. This relationship of the type $t_E=f(x,t_a), t'_E(x',t'_a)$ expresses the standard Lorentz transformations as a function of $t_a$ and $t'_a$ respectively. We present several cases of nonstandard Lorentz transformation (the case of radar detection, the case when one reference frame is filled with an ideal transparent dielectric and the case of relativity of the apparent, actual and synchronized positions of the same moving particle).*

## 1. Introduction
Leubner, Aufinger and Krumm[1] state that "In view of persisting confusion in the literature, it is not surprise at all that the undergraduate course of special relativity passes very swiftly over the question of clock synchronization. So swiftly in fact, that the beginner is left with the impression that there exists a single and essentially unique clock synchronization procedure (namely the **standard** or **Einstein** procedure despite the qualifier **convention** used by the instructor. As a consequence, students (and also many teachers) erroneously identify the properties of space-time with the familiar form that certain synchronization dependent quantities assume under the standard synchronization (length contraction, time dilation)."
Stimulated by this statement the teacher starts to prepare his own lecture on the subject. Looking for references, he finds out that many approaches to the subject are based on notions like "base vectors"[1], "the wave equation in an anisotropic space"[2]. These concepts are not in an easy reach for beginners and they make little use of Einstein's special relativity theory. An experienced teacher knows that such approaches find little audience in a class of students who already know about Einstein's special relativity and who consider that they are obliged to learn again the subject. He also likes to avoid using the concept of "two-way speed of light", a concept used by some other approaches to the subject.[3].

Consider the events E($x,y,z,t_E$) and E'($x',y',z',t'_E$) detected from the inertial reference frames I and I' in the standard arrangement. The events are characterized by their Cartesian space coordinates ($x,y,z$) and ($x',y',z'$) and by their time coordinates $t_E$ and $t'_E$ respectively. If the two events take place at the same point in space and the corresponding time coordinates represent the readings of two clocks C($x,y,z$) and C'($x',y',z'$) located at the point where the event takes place, each synchronized in its rest frame with the other clocks of that reference frame following a synchronization procedure proposed by Einstein[6], then the corresponding space-time coordinates are related by the **standard Lorentz-Einstein transformations**

$$x = \frac{x' + Vt'_E}{\sqrt{1 - \frac{V^2}{c^2}}} \qquad (1)$$

$$t_E = \frac{t'_E + \frac{V}{c^2}x'}{\sqrt{1 - \frac{V^2}{c^2}}} \qquad (2)$$

$$y = y' \qquad (3)$$
$$z = z' \qquad (4)$$

The inverse Lorentz-Einstein transformations are obtained by simply changing the corresponding unprimed physical quantities with primed ones and changing the sign of $V$ - the relative velocity of I and I'. The standard Lorentz-Einstein transformations are the consequences of the two postulates of Einstein's special relativity theory:
  (I) **The principle of relativity**: asserting the existence and the equivalence of inertial reference frames.
  (II) **The principle of the constancy of the one-way speed of light**: asserting the constancy of the one-way speed of light in empty space relative to these frames. Otherwise stated the light propagates in all directions of empty space with the same speed c.

In order to obtain a nonstandard Lorentz transformation we establish a physically correct relationship between the reading of the standard synchronized clock and the reading of another clock at rest in the same inertial reference frame expressing the Lorentz transformation as a function of the reading of the later clock. Following this strategy we have presented recently a simple approach to the consequences of Leubner's everyday clock synchronization.[1] The aim of this paper is to present several non-standard Lorentz transformations attractive by their simple and transparent shape.

## 2. Nonstandard Lorentz-Einstein transformations
### 2.1. Radar detection of the space-time coordinates of an event

The scenario we follow is sketched in Figure 1. It is detected from the inertial reference frame I' in a two-space dimensions approach. It involves the standard synchronized clocks of the I' frame C'(x'y') located at the different points of the plane defined by the axes of I' and clock $C'_0(0,0)$ located at the origin O', all displaying the same running time.

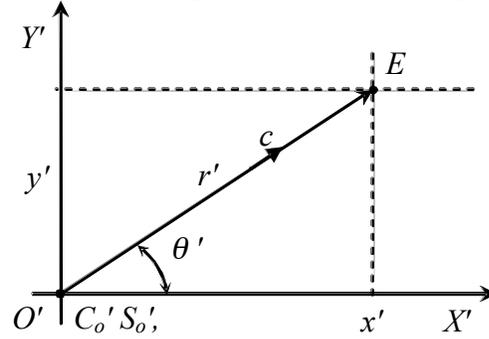

**Figure 1**. *Scenario for the radar detection of an event as detected from I'.*

When clock $C'_0$ reads $t'_e$, the source of light $S'_0(0,0)$ located at the origin O' emits light signals in all directions. One of the light signals emitted along a direction that makes an angle $\theta'$ with the positive direction of the O'X' axis arrives at the location of the clock C' the position of which is defined by the Cartesian space coordinates (x'y') or by the polar coordinates $(r',\theta')$, $r'$ representing the length of the position vector and $\theta'$ the polar angle. Arriving at clock C', the light signal generates the event $E'(x' = r'\cos\theta'; y' = r'\sin\theta'; t'_E = t'_e + r'/c)$ characterized by the space coordinates $x'=r'\cos\theta'$; $y'=r'\sin\theta'$ and by the time coordinate $t'_E = t'_e + r'/c$. Detected from the inertial reference frame I, the same event is characterized by the space-coordinates:

$$x = \frac{x' + Vt'_E}{\sqrt{1 - \frac{V^2}{c^2}}} = r' \frac{(\cos\theta' + V/c) + \dfrac{V/c}{r'/c} t'_e}{\sqrt{1 - \dfrac{V^2}{c^2}}} \qquad (5)$$

$$y = y' = r'\sin\theta' \qquad (6)$$

the lengths of the position vectors transforming as:

$$r = \sqrt{x^2 + y^2} = r'\sqrt{\sin^2\theta' + \frac{[(\cos\theta' + V/c) + \frac{V/c}{r'/c}t'_e]^2}{1 - V^2/c^2}} \qquad (7)$$

The transformation equations derived above are a typical example of **nonstandard Lorentz-Einstein transformations** which enable observers from I to express the length of the position vector $r$ detected from I as a function of its length detected from I' and of the running time $t'_e$ displayed by the clock $C'_0$ when the radar signal is emitted and not as a function of the time coordinate $t'_E$ displayed by the clock located where the signal is received. If when detected from I' the radar signal propagates along a direction $\theta'$ then when detected from I it propagates along a direction $\theta$, the two angles being related by:

$$\tan\theta = \frac{y}{x} = \frac{\sqrt{1 - \frac{V^2}{c^2}}\sin\theta'}{\cos\theta' + \frac{V}{c} + \frac{V/c}{r'/c}t'_e}. \qquad (8)$$

The time coordinates of the involved events transform as

$$t_E = \frac{t'_E + \frac{V}{c^2}x'}{\sqrt{1 - \frac{V^2}{c^2}}} = \frac{t'_e + \frac{r'}{c}(1 + \frac{V}{c}\cos\theta')}{\sqrt{1 - \frac{V^2}{c^2}}}. \qquad (9)$$

Because

$$t_E = t_e + \frac{r}{c} \qquad (10)$$

we obtain that the emission times transform as

$$t_e = \frac{t'_e + \frac{r'}{c}(1 + \frac{V}{c}\cos\theta')}{\sqrt{1 - \frac{V^2}{c^2}}} - \frac{r'}{c}\sqrt{\sin^2\theta' + \frac{(\cos\theta' + \frac{V}{c} + \frac{V/c}{r'/c})}{1 - \frac{V^2}{c^2}}}. \qquad (11)$$

In the particular case when $t'_e = 0$ the transformation equations (7) and (9) derived above become

$$r = r'\frac{1 + \frac{V}{c}\cos\theta'}{\sqrt{1 - \frac{V^2}{c^2}}} \qquad (12)$$

and

$$t_E = t'_E \frac{1 + \frac{V}{c}\cos\theta'}{\sqrt{1 - \frac{V^2}{c^2}}} = Dt'_E \tag{13}$$

the transformation being performed via the Doppler factor D defined as

$$D = \frac{1 + \frac{V}{c}\cos\theta'}{\sqrt{1 - \frac{V^2}{c^2}}}. \tag{14}$$

**2.2. Lorentz transformations for the case when the inertial reference frame I' is filled with an ideal transparent dielectric at rest relative to it.**
Consider that the inertial reference frame I' is at rest in a transparent perfect dielectric characterized by its proper refractive index $n > 1$. A light signal propagates there along the positive direction of the O'X' axis with speed

$$c_n = \frac{c}{n}. \tag{15}$$

Consider that at a point M'(x') located on the O'X' axis we find the clocks $C'_1(x')$ and $C'_2(x')$. The first one is synchronized with clock $C'_0(0)$ located at the origin O' following Einstein's clock synchronization procedure and we suppose that it displays a time:

$$t'_E = \frac{x'}{c}. \tag{16}$$

Clock $C'_x(x')$ is synchronized with clock $C'_0(0)$ using a synchronizing signal that propagates with speed $c_n = c/n$ displaying a time

$$t'_n = \frac{nx'}{c}. \tag{17}$$

Combining (16) and (17) we obtain

$$t'_E = t'_n + \frac{x'}{c}(1-n). \tag{18}$$

Performing the Lorentz transformations to the inertial reference frame I the result is:

$$x = \frac{x'[1 + \frac{V}{c}(1-n)]}{\sqrt{1 - \frac{V^2}{c^2}}} + \frac{Vt'_n}{\sqrt{1 - \frac{V^2}{c^2}}} \tag{19}$$

$$t_E = \frac{t'_n}{\sqrt{1 - \frac{V^2}{c^2}}} + \frac{\frac{x'}{c}(1 - n + \frac{V}{c})}{\sqrt{1 - \frac{V^2}{c^2}}}. \tag{20}$$

Now choose *n* to be the refraction index that makes equation (20) independent of *x'* i.e.
$$1 - n + \frac{V}{c} = 0 \tag{21}$$
which solves as
$$n = 1 + \frac{V}{c} \tag{22}$$
and so, in that particular case
$$c_n = \frac{c}{1 + \frac{V}{c}}. \tag{23}$$
Substitution in (19) and (20) gives
$$x = x'\sqrt{1 - \frac{V^2}{c^2}} + \frac{Vt'_n}{\sqrt{1 - \frac{V^2}{c^2}}} \tag{24}$$
and
$$t_E = \frac{t'_n}{\sqrt{1 - \frac{V^2}{c^2}}}. \tag{25}$$
The inverse transformations obtained by combining (24) and (25) are
$$x' = \frac{x - Vt_E}{\sqrt{1 - \frac{V^2}{c^2}}} \tag{26}$$
$$t'_n = t_E \sqrt{1 - \frac{V^2}{c^2}}. \tag{27}$$
The speeds $u_E = \frac{x}{t_E}$ and $u'_n = \frac{x}{t_E}$ are related by
$$u'_n = \frac{u_E - V}{1 - \frac{V^2}{c^2}} \tag{28}$$
the inverse transformation being
$$u_E = u'_n(1 - \frac{V^2}{c^2}) + V. \tag{29}$$
We have recovered that way the results obtained by Selleri[4], Abreu and Guerra[5]. Selleri[4] derives the equations we have obtained above starting with the following assumptions:
(i) Space is homogeneous and isotropic and time homogeneous, at list if judged by observers at rest in I:

(ii) In the *isotropic* system I the velocity of light is "c" in all directions, so that clocks can be synchronized in I and one-way velocities relative to I can be measured;
(iii) The origin of I', observed from I, is seen to move with velocity V<c parallel to the +x axis, that is according to the equation x=Vt;
(iv) The axes of I and I' coincide for t=t'=0;
(v) The two-way velocity of light is the same in all directions and in all inertial systems;
(vi) Clock retardation takes place with the usual velocity dependent factor when clocks move with respect to the isotropic reference frame I.

The fact that our approach presented above and the one proposed by Selleri lead to the same result is due to the fact that I' is filled in our case with a transparent dielectric and its clocks are being synchronized by a signal propagating with speed $c_n=c/n$. Abreu and Guerra[5] and obtain similar results due to the fact that they derive equation (26) as a result of the length contraction effect and equation (27) as a result of the so called external clock synchronization.

## 2.3. Special relativity and the apparent, actual and synchronized positions of the same particle, moving with constant speed V relative to I in the positive direction of the OX axis.

In order to keep the problem as simple as possible consider the one space dimensions scenario sketched in Figure 2 as detected from I.

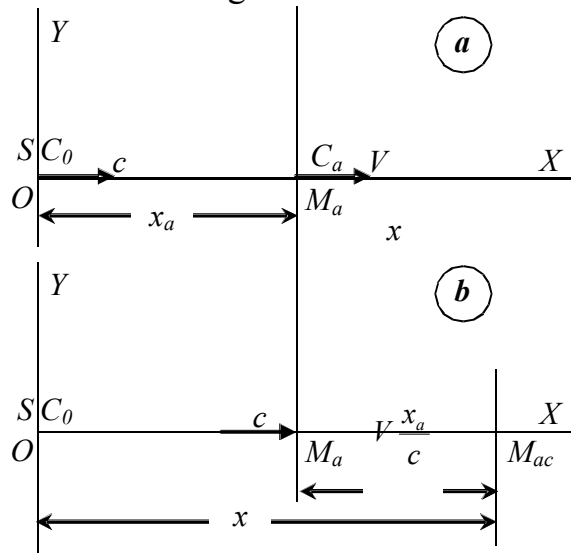

*Figure 2. Scenario for the definition of apparent position $M_a$ and of the actual position of the same particle moving with constant speed V in the positive direction of the overlapped OX axis.*

In Figure 2a $M_a(x_a)$ represents the location of a particle that moves with speed $V$ in the positive direction of the OX axis when the standard synchronized clocks of I read t=0. At same time a source of light S(0) located at the origin O in frame I emits a light signal in the positive direction of the OX axis. We call $M_a(x_a)$ the **apparent position** of the moving particle. In Figure 2b M(x) represents the location of the same particle when the light signal arrives at point $M_a$. M(x) represents what we call the **actual position**. The coordinates $x_a$ and x are related by

$$x = x_a + \frac{V}{c}x_a \tag{30}$$

The event $E_a(x_a, t_a = x_a/c)$ is associated with the arrival of the light signal at the apparent position whereas the event $E(x = x_a(1+V/c), t_a = x_a/c)$ is associated with the arrival of the particle at the actual position and the arrival of the light signal at the apparent position, the two events being simultaneous but taking place at different points in space.

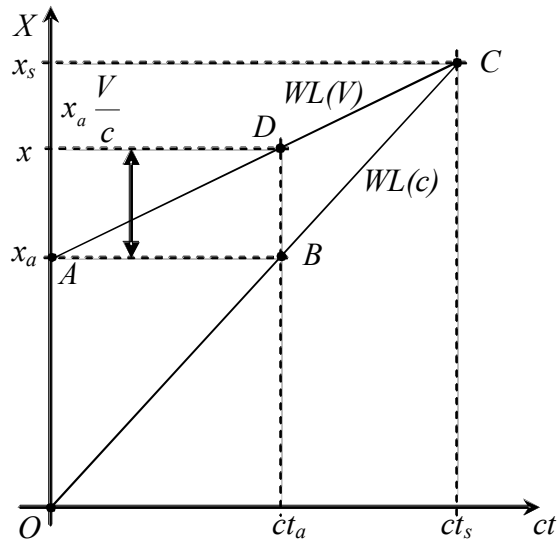

*Figure 3. A classical space-time diagram displays the events associated with the apparent position B, with the actual position D and with the synchronized position C.*

Figure 3 illustrates the scenario we propose on a **classical** space-time diagram on the axes of which we measure the space coordinates $x$ and the product $ct$ between the speed of light in empty space and respectively the time coordinate of the involved events. WL(c) represents the world line of the light signal emitted at t=0 from the origin O whereas WL(V) represents the world line of the particle moving with speed $V$ described by (30). $M_a$ represents the event associated with the apparent position whereas M represents the event associated with the actual position. Performing the Lorentz transformations of the space-time coordinates of event E to the rest frame I' of the moving particle we obtain:

$$x' = x_a [\sqrt{\frac{1+\frac{V}{c}}{1-\frac{V}{c}}} - \frac{\frac{V}{c}}{\sqrt{1-\frac{V^2}{c^2}}}] \tag{31}$$

and

$$t'_E = t_a [\frac{1}{\sqrt{1-\frac{V^2}{c^2}}} - \frac{V}{c}\sqrt{\frac{1+\frac{V}{c}}{1-\frac{V}{c}}}] . \tag{32}$$

The intersection of the world lines WL(c) and WL(V) generates the event $E_s$ characterized by the space coordinate $x_s$ obtained from (30)

$$x_s = x_a + \frac{V}{c} x_s \tag{33}$$

i.e.

$$x_s = \frac{x_a}{1-\frac{V}{c}} . \tag{34}$$

Event $E_s$ is characterized by a time coordinate $t_s = x_s/c$. Performing the Lorentz transformations to the rest frame of the moving particle I' we obtain the nonstandard Lorentz transformations

$$x'_s = \frac{x_a}{\sqrt{1-\frac{V^2}{c^2}}} . \tag{34}$$

$$t'_E = \frac{t_a}{\sqrt{1-\frac{V^2}{c^2}}} .$$

resulting that

$$u'_E = u_s . \tag{35}$$

$u'_E$ and $u_s$ representing the speeds of the same particle measured in I' and in I respectively.
The location $M_s$ of the moving particle represents in this case what we call the **synchronized** position.

## 3. Conclusions

The important contribution of our paper consists in the fact that it shows that the standard Lorentz transformations account for the consequences of any arbitrary but "physically" correct clock synchronization procedures. Moreover they lead to the corresponding transformation equations for the space-time coordinates of the involved events.

## References

[1]C.Leubner ,K.Aufinger and P.Krumm, "Elementary relativity with "everyday" clock synchronization", Eur.J.Phys. **13**, 170-177 (1992)

[2]Abraham A.Ungar, "Formalism to deal with Reichenbach's special theory of relativity", Foundation of Physics, **21,** 691-724 (1991)

[3]H.Reichenbach, *Axiomatization of the Theory of Relativity, trans.and ed.by M.Reichenbach,* University of California Press, Berkeley (1969)

[4]F.Selleri, "Non-invariant one-way speed of light", Foundations of Physics, **26,** 641-664 (1996)

[5]Rodrigo de Abreu and Vasco Guerra, *Relativity :Einstein's lost frame*, ]Extra[muros 2005 and references therein.